\documentclass[aps,pra,a4paper,twocolumn,showpacs]{revtex4}
\usepackage[dvips]{epsfig}
\usepackage[usenames]{color}
\usepackage{pstricks}   
\usepackage{amsmath}
\usepackage{amssymb}
\usepackage{subfigure}

\begin{document}

\title{Formation of molecular oxygen in ultracold O + OH reaction}

\author{Goulven Qu{\'e}m{\'e}ner, Naduvalath Balakrishnan}
\affiliation{Department of Chemistry, University of Nevada Las Vegas,
Las Vegas, NV-89154, USA}

\author{Brian K. Kendrick}
\affiliation{Theoretical Division, Los Alamos National Laboratory,
Los Alamos, NM-87545, USA}

\date{\today}

\begin{abstract}
We discuss the formation of molecular oxygen  in
ultracold collisions between hydroxyl radicals and atomic oxygen. 
A time-independent quantum formalism based on hyperspherical coordinates is
employed for the calculations.
Elastic, inelastic and reactive cross sections
as well as the vibrational and rotational populations
of the product O$_2$ molecules are reported.
A $J$-shifting approximation is used to compute the 
rate coefficients.
At temperatures $T = 10 - 100$~mK for which the OH molecules have been cooled and trapped
experimentally,
the elastic and reactive rate coefficients 
are of comparable magnitude,
while at colder temperatures, $T < 1$~mK,
the formation of molecular oxygen becomes the dominant pathway.
The validity of a classical capture model to describe cold collisions of OH and O
is also discussed.
While very good agreement is found
between classical and quantum results at $T=0.3$~K,
at higher temperatures,
the quantum calculations predict a larger rate
coefficient than the classical model, in agreement with experimental data
for the O + OH reaction. The
zero-temperature limiting value of the rate coefficient 
is predicted to be about $6 \times 10^{-12}$~cm$^3$~molecule$^{-1}$~s$^{-1}$,
a value comparable to that of barrierless alkali-metal atom - dimer systems
and about a factor of five larger than that of the tunneling dominated F + H$_2$ reaction.
\end{abstract}

\pacs{ }

\maketitle

\font\smallfont=cmr7

\section{Introduction}

Important experimental progress
is being made
in creating ultracold molecules
in tightly bound vibrational levels~\cite{Gould08}.
Very recently, formation of ground state molecules in the 
vibrational level $v=0$ has been reported
by different groups
for homonuclear molecules such as Cs$_2$~\cite{Viteau08}, Rb$_2$~\cite{Lang08},
as well as for heteronuclear polar molecules such as RbCs~\cite{Sage05}, KRb~\cite{Ni08}, 
and LiCs~\cite{Deiglmayr08}.
There has also been much progress in the measurement of rate coefficients
of barrierless reactions involving alkali-metal atoms
at cold and ultracold temperatures.
This includes atom - molecule collisions
such as Rb + Rb$_2$~\cite{Wynar00,Syassen06},
Cs + Cs$_2$~\cite{Staanum06,Zahzam06},
Na + Na$_2$~\cite{Mukaiyama04},
Rb/Cs + RbCs~\cite{Hudson08},
and molecule - molecule collisions
such as Cs$_2$ + Cs$_2$~\cite{Zahzam06,Ferlaino08},
Na$_2$ + Na$_2$~\cite{Mukaiyama04}, and
Rb$_2$ + Rb$_2$~\cite{Syassen06}.
The typical rate coefficients of these reactions
are on the order of 
$10^{-11} - 10^{-10}$~cm$^3$~molecule$^{-1}$~s$^{-1}$
depending on the collisional system, the temperature and
the vibrational levels probed.
All these experimental studies indicate that 
inelastic and reactive processes
occur at significant rates at ultracold temperatures,
in accordance with theoretical predictions
on barrier reactions~\cite{Bala01,Bodo02,Bodo04,Weck05,Weck06,Quemener08b} as well
as a number of alkali-metal atom - dimer 
systems~\cite{Soldan02,Quemener04,Quemener05,Cvitas05a,Cvitas05b,Hutson06,Hutson07a,Quemener07,Cvitas07,Quemener08c}
such as Li + Li$_2$, Na + Na$_2$ and K + K$_2$ which
proceeds without an energy barrier in the entrance channel.
The alkali-metal systems are characterized by triatomic complexes 
with deep potential wells
which make them challenging systems for accurate quantum calculations.
For these heavy systems,
the density of triatomic states
is large and it leads to strong couplings
between them, enhancing 
inelastic quenching or reactive scattering~\cite{Quemener-CHAPTER}.
However, explicit quantum calculations of molecule - molecule systems involving 
alkali-metal atoms are computationally intractable. 
In  a recent theoretical study of
vibrational relaxation in collisions
between H$_2$ molecules, 
Qu{\'e}m{\'e}ner et al.~\cite{Quemener08a}
showed that the relaxation rate coefficients 
at ultralow temperature
can attain large values for some near-resonant processes 
which involve simultaneous conservation of internal energy
and total internal rotational angular momentum of the colliding molecules.

While there have been a number of theoretical studies of ultracold  reactive collisions
of tunneling dominated reactions 
with chemically distinct reactants and 
products~\cite{Bala01,Bodo02,Bodo04,Weck05,Weck06,Quemener08b},
there have been no
such studies involving barrierless chemical reactions
at ultracold temperatures.
Here we investigate the exothermic 
reaction
\begin{eqnarray}
\text{O$(^3\text{P})$ + OH$(^2\Pi)$ $\to$ H$(^2\text{S})$ + O$_2(^3\Sigma_g^-)$}
\label{OpOH}
\end{eqnarray}
at cold and ultracold temperatures
as an example of a barrierless chemical reaction involving non-alkali-metal atom
systems.
The reaction is of key interest in oxygen chemistry in the interstellar medium,
OH chemistry in the upper stratosphere and mesosphere, 
and combustion chemistry (see Ref.~\cite{Quemener08d}
and references therein).
The OH radical has also been cooled and trapped using the buffer gas 
and stark decelerator techniques~\cite{Bochinski03,Bochinski04,Meerakker05a,Sawyer07}.
High precision measurements of 
its radiative lifetime~\cite{Meerakker05b}
and its hyperfine constant~\cite{Hudson06a,Lev06} have recently
been reported.
An experimental study of scattering between cold OH molecules and He atoms and D$_2$ molecules
has  recently been reported~\cite{Sawyer08}. 
The cooling and trapping studies have stimulated a number of theoretical
investigations involving the OH molecule in the last few years.
Gonz{\'a}lez--S{\'a}nchez et al.~\cite{Gonzalez06}
reported rotational relaxation and spin-flipping 
of OH in He collisions at ultralow energies.
They found that rotational relaxation occurs more efficiently 
than  elastic collisions
at vanishing collision energies.
By carrying out quantum calculations of Rb + OH collisions at ultracold 
temperatures Lara et al.~\cite{Lara06,Lara07} explored 
the possibility of sympathetic cooling of OH molecules by collisions with 
Rb atoms.
They argued that efficient
sympathetic cooling of OH molecules in the ground vibrational state by collisions with Rb
atoms is 
unlikely to occur due to the large inelastic rate coefficient.
External fields can also have important effects
on ultracold molecular collisions~\cite{KremsIRPC,KremsPCCP,Tscherbul08}.
Avdeenkov and Bohn
investigated
the effect of external fields
on ultracold collisions between OH 
or OD molecules~\cite{Avdeenkov02,Avdeenkov03,Avdeenkov05}.
Ticknor and Bohn~\cite{Ticknor05} also studied
OH$-$OH collisions
in the presence of a magnetic field.
They showed that magnetic fields of several thousand Gauss
suppress inelastic collision rates
by about two orders of magnitude.
In a recent study~\cite{Quemener08d}, 
we reported 
quantum dynamics calculations
of reaction~\eqref{OpOH} for $T=10 - 600$~K and
found no significant
decrease of the rate coefficient
in the temperature 
range  $39 - 10$~K, in accordance with conclusions of a recent
experimental study by Carty et al.~\cite{Carty06}. 
Our calculations for reaction probabilities
were in excellent agreement with those of Xu et al.~\cite{Xu07}
for collision energies $E_c > 0.012$~eV.

In this paper, we present the quantum dynamics
of reaction~\eqref{OpOH} 
at low and ultralow collision energies to explore the
behavior of complex forming chemical reactions 
at cold and ultracold temperatures.
Since OH molecules have been experimentally
cooled and trapped at low temperatures, we believe that
collisional properties of the O + OH reaction will be of considerable interest to
the cooling and trapping community.
The paper is organized as follows: The details of the quantum mechanical formalism
along with convergence tests are discussed in
Section II. Results of
cross sections, rate coefficients, and state-to-state product distributions
are given in Section III. 
We also include in this section a  discussion on the usefulness of a classical model in describing cold collisions of O and OH.
Conclusions are presented in Section IV.

\section{Methodology}

\subsection{Potential energy surfaces}

\begin{figure} [h]
\begin{center}
\includegraphics*[width=8cm,keepaspectratio=true,angle=0]{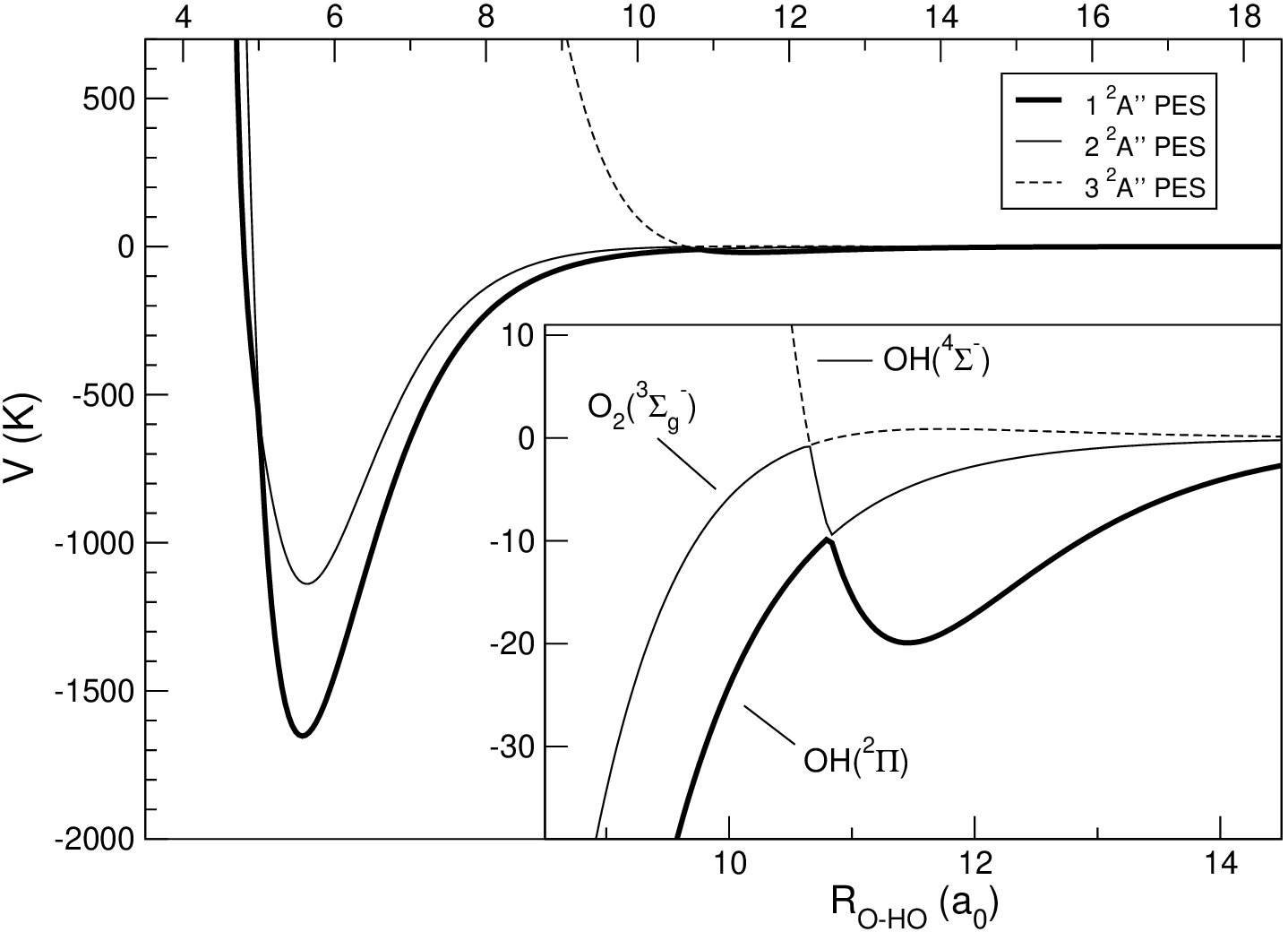}
\caption{
Potential energy curves for the 3 lowest $^2A''$ states of HO$_2$ for the DIMKP PES,
for a O--HO linear approach with $r_{\text{OH}}=1.83 \ a_0$.
The conical intersections arising at $R_{\text{O-HO}} \approx 10.8 \ a_0$ 
is shown in the inset. 
\label{CURVE-FIG}
}
\end{center}
\end{figure}

We employed a modified version of 
the electronically adiabatic ground state $(1 \ ^2A'')$ 
potential energy surface (PES) of HO$_2$
calculated by Kendrick and Pack~\cite{Kendrick95} 
using a diatomics-in-molecule (DIM) formalism.
This new version includes improvements to the long-range behavior
and is referred to as the DIMKP PES. 
In particular, a switching function was implemented which more smoothly ``turns-on" 
the long-range van der Waals potentials for the electronic ground states
of both O$_2$ and OH. This switching function is given by
$f_{\rm switch} = 0.5({\rm tanh}(\alpha(r - r_0)) + 1)$ where $\alpha=1$,
$r_0=7.0 \ a_0$ for O$_2(^3\Sigma^-_g)$ and $r_0 = 10 \ a_0$ for OH($^2\Pi$).
A minor global refitting of the DIM HO$_2$ PES
to the original set of {\it ab initio} data was required in order to
account for the new switching functions and ensure a smooth transition 
to the long-range behavior. The same long-range coefficients were used as
in the original version of the surface, 
for OH($^2\Pi$): 
$C_6^{\text{O--H}}=  9.295 \ E_h a_0^6$,
$C_8^{\text{O--H}} = 169.09 \ E_h a_0^8$, 
$C_{10}^{\text{O--H}} = 4060.85 \ E_h a_0^{10}$, 
and for O$_2(^3\Sigma^-_g)$:
$C_6^{\text{O--O}}  = 14.89 \ E_h a_0^6$, 
$C_8^{\text{O--O}} = 206.67 \ E_h a_0^8$, and 
$C_{10}^{\text{O--O}} = 3753.745 \ E_h a_0^{10}$~\cite{Kendrick95}.
The global fit in the interaction region is essentially identical to the 
original fit with nearly the same rms deviation of 0.099~eV (2.3~kcal/mol).
The improvements to the long-range behavior are important for the 
ultracold collisions studied in this work but should not significantly affect the
results of previous scattering calculations at higher (thermal)
energies~\cite{Kendrick96}. 
We present in Fig.~\ref{CURVE-FIG} the potential energy curves 
for the 3 lowest $^2A''$ states of HO$_2$ for the DIMKP PES,
for a O--HO linear approach.
We also note in the inset of Fig.~\ref{CURVE-FIG} that the DIMKP PES predicts a shallow 
conical intersection along the O--HO approach due to the crossing of the
OH($\Pi$) and OH($\Sigma$) states. For a fixed $r_{\text{OH}}=1.83 \ a_0$, 
this crossing occurs at $R_{\text{O-HO}} = 10.8235 \ a_0$ and its energy lies
$8.30 \times 10^{-4}$~eV ($\approx 9.6$~K) {\it below} the asymptotic energy of the
O + OH channel.
For comparison purposes,
we also employed the {\it ab initio} PES computed by
Xu, Xie, Zhang, Lin, and Guo~\cite{Xu05,Xie07}, referred to as the XXZLG PES.
The XXZLG PES has been used in a number of quantum dynamics calculations of the
O + OH system at high collision 
energies~\cite{Xu07,Lin08,Quan08}. 
The present DIMKP PES was employed
for the first time
for this reaction
in our previous work~\cite{Quemener08d}
and it is preferred
at low energies as it includes accurate long-range coefficients.

\subsection{Quantum mechanical approach and convergence tests}

The quantum dynamics calculations have been performed using the
adiabatically adjusting principle-axis hyperspherical (APH) approach
of Pack and Parker~\cite{Pack87}. The method uses the democratic 
Smith-Whitten hyperspherical coordinates in the inner 
region 
($\rho < 17 \ a_0$)
that includes the triatomic well of the HO$_2$ system
and the Delves--Fock hyperspherical coordinates in the outer region
($\rho > 17 \ a_0$)
in the valleys of the H + O$_2$ and O + OH arrangement channels.
For a given value of the total angular momentum quantum number, $J$, 
and  hyperspherical radius, $\rho$,
the wavefunction is expanded onto a basis set of adiabatic functions, which are eigenfunctions 
of a triatomic hyperangular Hamiltonian.
A hybrid DVR/FBR primitive basis set~\cite{Kendrick99} combined with
an Implicity Restarted Lanczos algorithm 
is used to diagonalize the hyperangular Hamiltonian matrix.
The time-independent Schr{\"o}dinger equation
yields a set of differential close-coupling equations in $\rho$, which are solved using
the log-derivative matrix propagation method of Johnson~\cite{Johnson73}.
The log-derivative matrix is propagated
to a matching distance 
($\rho_m$)
where asymptotic boundary conditions 
are applied to evaluate the reactance matrix $K^{J}$ and the scattering matrix $S^{J}$. 
The square elements of the $S^{J}$ matrix provide
the state-to-state transition probabilities, $P^{J}$. The matching distance and all other
parameters employed in the calculations were determined by optimization and
extensive convergence studies.
To secure convergence of the reaction probabilities,
 the number of hyperspherical channels, $n$, included
in the close coupling equations is 393. This is  sufficient
to obtain converged results at low and ultralow energies.
The
cross sections are calculated using the standard formulae:
\begin{align*}
\sigma^J_{\text{el}} &= \frac{\pi}{\text{k}^2} |1-S^J_{ii}|^2
& \sigma^J_{\text{re}} &=
\frac{\pi}{\text{k}^2} \sum^{}_{\text{reactive $j$}}|S^J_{ij}|^2,
\end{align*}
where $i,j$ denote initial and final quantum states.
The non-thermal elastic and reactive rate coefficients are given by $\sigma_{\text{el}} \times \nu$
and $\sigma_{\text{re}} \times \nu$ 
where 
$\nu = \hbar \text{k}/\mu$ is the incident velocity for relative motion
of the O atom and the OH molecule.

\begin{table}[h]
\begin{center}
\begin{tabular}{c c c c c c}
\hline
$\rho_m$ ($a_0$) & 26.8 & 32.7 & 39.9 & 44.0   \\ [0.5ex]
\hline
$E_c$ (eV) & $\sigma^{J=0}_{\text{el}}$ & & & \\
10$^{-10}$ & 0.4995 & 0.5019 & 0.4786 & 0.4679  \\
10$^{-9}$ & 0.4965  & 0.4977 & 0.4742 & 0.4635 \\
10$^{-8}$ & 0.4877  & 0.4853 & 0.4610  & 0.4505  \\
10$^{-7}$ & 0.4596 & 0.4472 &  0.4206 &  0.4110 \\
10$^{-6}$ & 0.3708 & 0.3369 & 0.3085  & 0.3021  \\
10$^{-5}$ & 0.1432  & 0.1164 & 0.1108  & 0.1123  \\
10$^{-4}$ & 0.005536 & 0.005214 &  0.005300 & 0.005217  \\
10$^{-3}$ & 0.001531  & 0.001516 & 0.001506  & 0.001507  \\ [0.5ex]
\hline
\end{tabular} 
\end{center}
\caption{Convergence of the elastic cross section $\sigma^{J=0}_{\text{el}}$
in units of 10$^{-13}$cm$^2$
with the matching distance $\rho_m$
for O + OH($v=0,j=0$)
using the DIMKP PES, for $n=393$.
\label{TAB1}
}
\end{table}

\begin{table}[h]
\begin{center}
\begin{tabular}{c c c c c c}
\hline
$\rho_m$ ($a_0$) & 26.8 & 32.7 & 39.9 & 44.0   \\ [0.5ex]
\hline
$E_c$ (eV) & $\sigma^{J=0}_{\text{re}} \times \nu$ & & & \\
10$^{-10}$ & 0.1965 & 0.2777 & 0.3123 & 0.3133 \\
10$^{-9}$ & 0.1954 & 0.2755 & 0.3095 & 0.3104 \\
10$^{-8}$ & 0.1920 & 0.2688 & 0.3010 &  0.3018 \\
10$^{-7}$ & 0.1819 & 0.2490 & 0.2760  &  0.2765 \\
10$^{-6}$ & 0.1544 & 0.1979 & 0.2123 & 0.2119 \\
10$^{-5}$ & 0.09835 & 0.1080 & 0.1080 &  0.1073\\
10$^{-4}$ & 0.03748 & 0.03713 &  0.03724 &  0.03714 \\
10$^{-3}$ & 0.008704  & 0.008745  & 0.008761  & 0.008765 \\ [0.5ex]
\hline
\end{tabular}
\end{center}
\caption{Convergence of the non-thermal
reactive rate coefficients, $\sigma^{J=0}_{\text{re}} \times \nu$,
in units of $10^{-10}$~cm$^3$~molecule$^{-1}$~s$^{-1}$
with the matching distance $\rho_m$
for O + OH($v=0,j=0$)
using the DIMKP PES, for $n=393$.
\label{TAB2}
}
\end{table}

The convergence of 
the elastic cross section $\sigma_{\text{el}}^{J=0}$
and the non-thermal reactive rate coefficient $\sigma_{\text{re}}^{J=0} \times \nu$ 
are presented in Tab.~\ref{TAB1} and Tab.~\ref{TAB2} for O + OH($v=0,j=0$) on the DIMKP PES.
At vanishing  collision energies, these quantities attain finite 
values  as required by
the Bethe--Wigner laws~\cite{Bethe35,Wigner48}:
\begin{align*}
\sigma^{J=0}_{\text el} &\to \text{const.}
& \sigma^{J=0}_{\text{re}} \times \nu &\to \text{const.} 
\end{align*}
The tables also show convergence of the results with the matching distances.
At low energies the elastic cross section  converges less rapidly with the matching
distance $\rho_m$ than the reactive one.
This is because  the long-range contribution 
to the interaction potential is not negligible compared to the kinetic energy in 
the entrance channel, even for moderately large values of the hyperradius. Thus,
elastic cross sections are generally more
sensitive to the long-range tail of the interaction potential. 
For the parameters in Tab.~\ref{TAB1} and Tab.~\ref{TAB2} the  reactive rate coefficients
are converged to within  1\%
while the elastic cross sections are converged within 3\%.
All of the final results presented here are obtained using $\rho_m = 44.0$~a$_0$, which
is especially necessary to get converged results at $E_c < 10^{-4}$~eV~\cite{NOTE}.

\section{Results and discussion}

\subsection{Cross sections and rate coefficients}

\begin{figure} [t]
\begin{center}
\includegraphics*[width=8cm,keepaspectratio=true,angle=0]{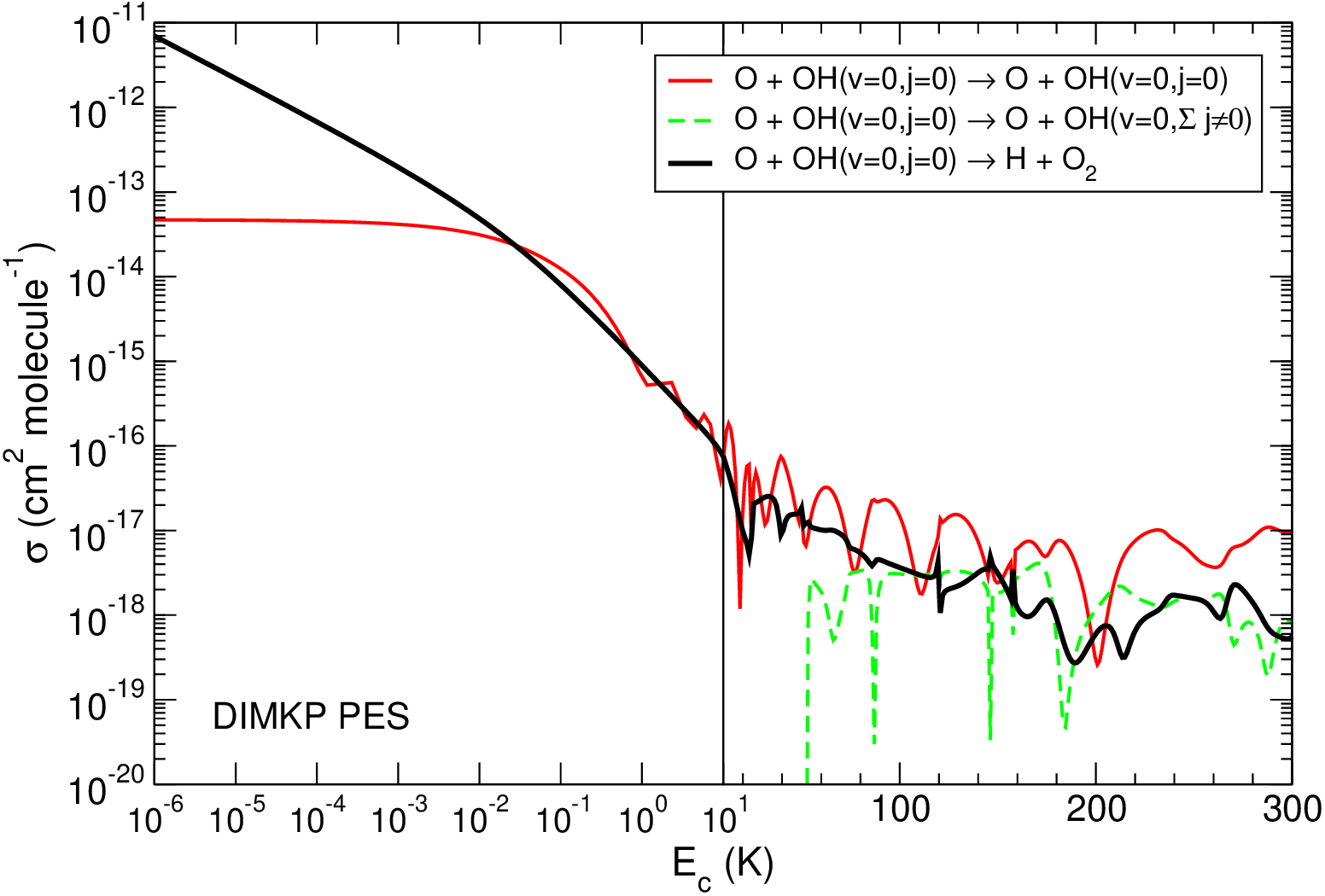}
\caption{
(Color online)
$J=0$ cross section
as a function of the collision energy
for the elastic (red curve), inelastic (green dashed curve) and reactive (black curve)
collisions of O atoms with OH$(v=0,j=0)$ molecules for the DIMKP PES.
\label{XS-DIMKP-FIG}
}
\end{center}
\end{figure}
\begin{figure} [h]
\begin{center}
\includegraphics*[width=8cm,keepaspectratio=true,angle=0]{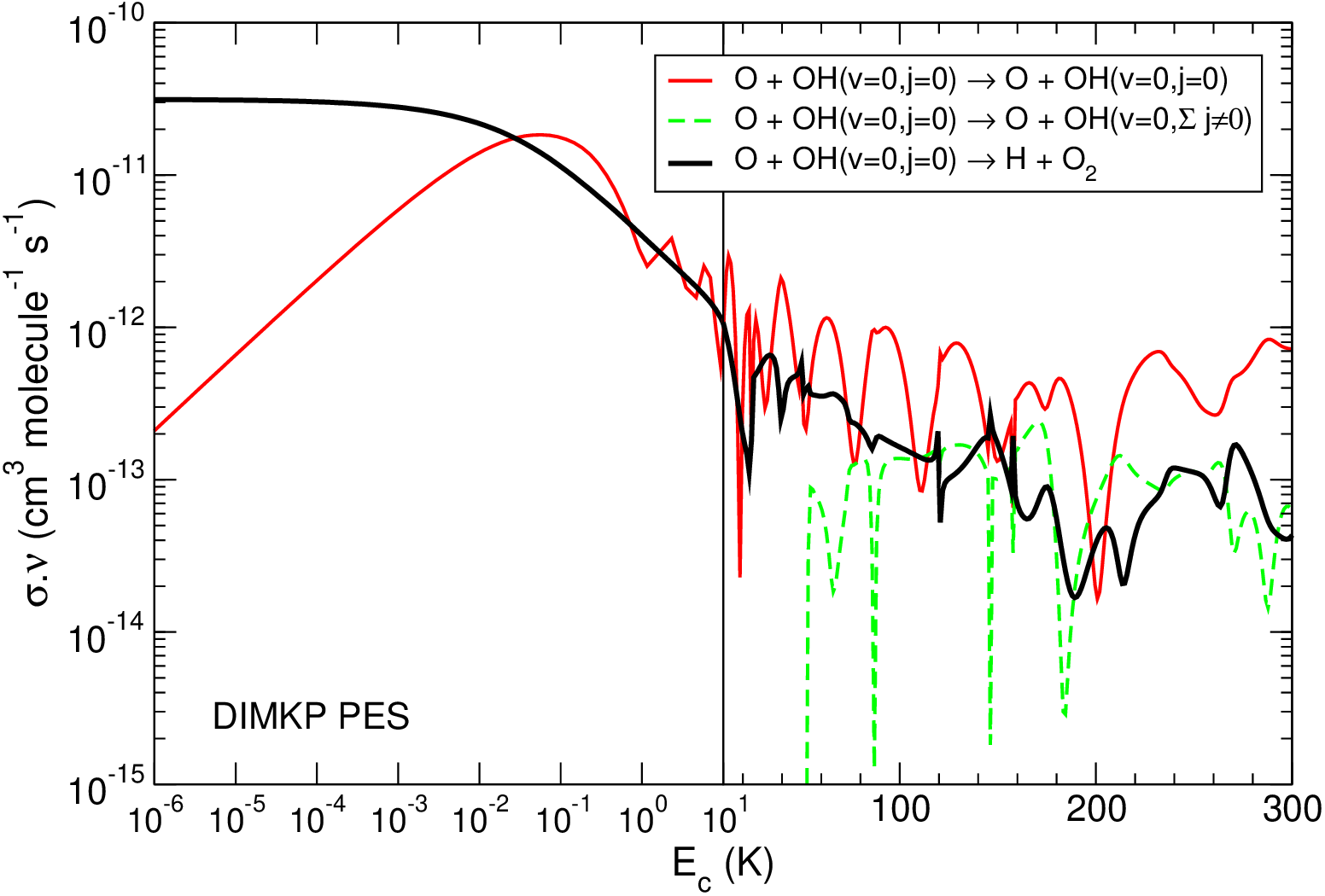}
\caption{
(Color online)
$J=0$ non-thermal rate coefficient
as a function of the collision energy
for the elastic (red curve), inelastic (green dashed curve) and reactive (black curve)
collisions of O atoms with OH$(v=0,j=0)$ molecules for the DIMKP PES.
\label{RATE-DIMKP-FIG}
}
\end{center}
\end{figure}
\begin{figure} [h]
\begin{center}
\includegraphics*[width=8cm,keepaspectratio=true,angle=0]{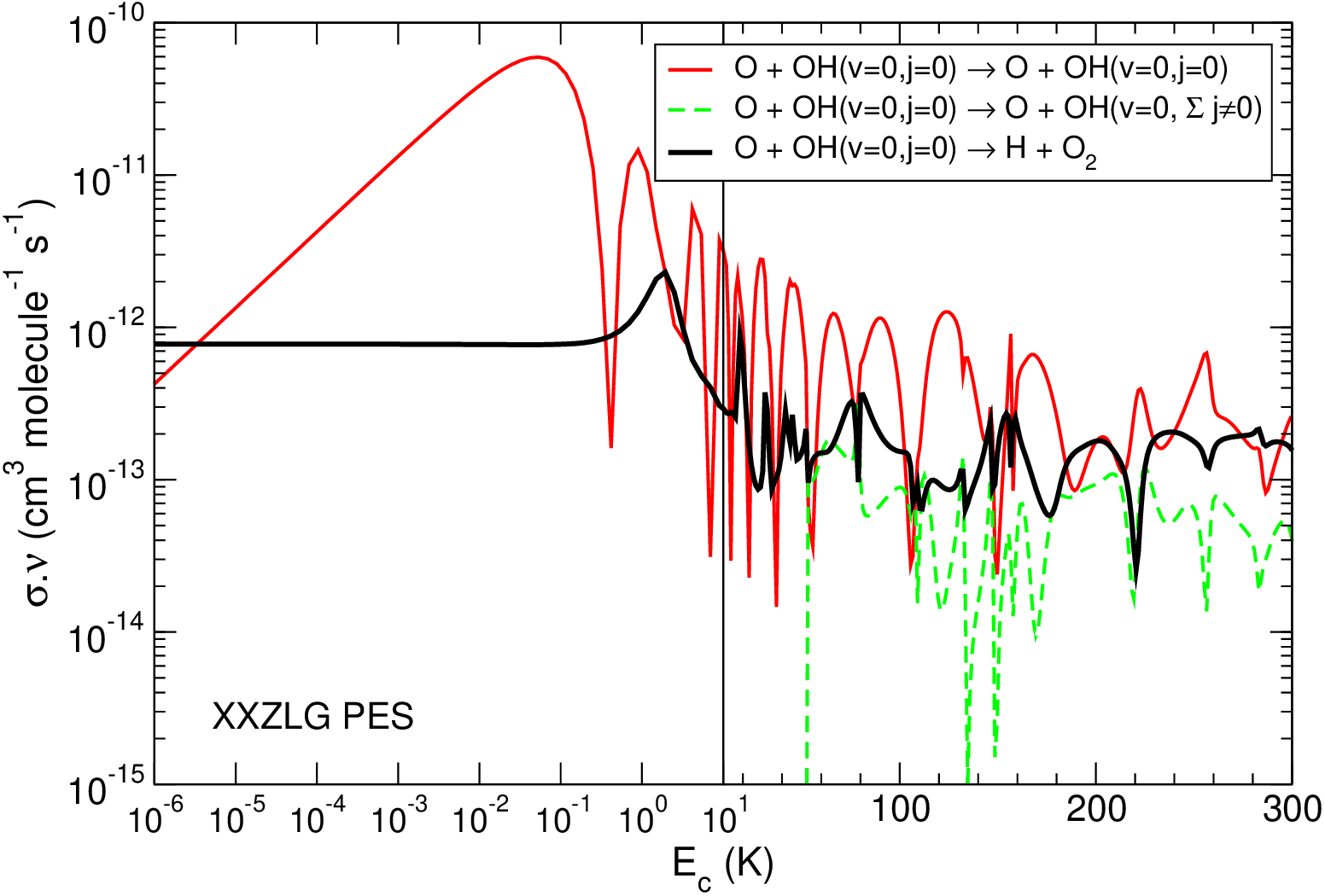}
\caption{
(Color online)
Same as Fig.~\ref{RATE-DIMKP-FIG}
but for the XXZLG PES.
\label{RATE-XXZLG-FIG}
}
\end{center}
\end{figure}
\begin{figure} [h]
\begin{center}
\includegraphics*[width=8cm,keepaspectratio=true,angle=0]{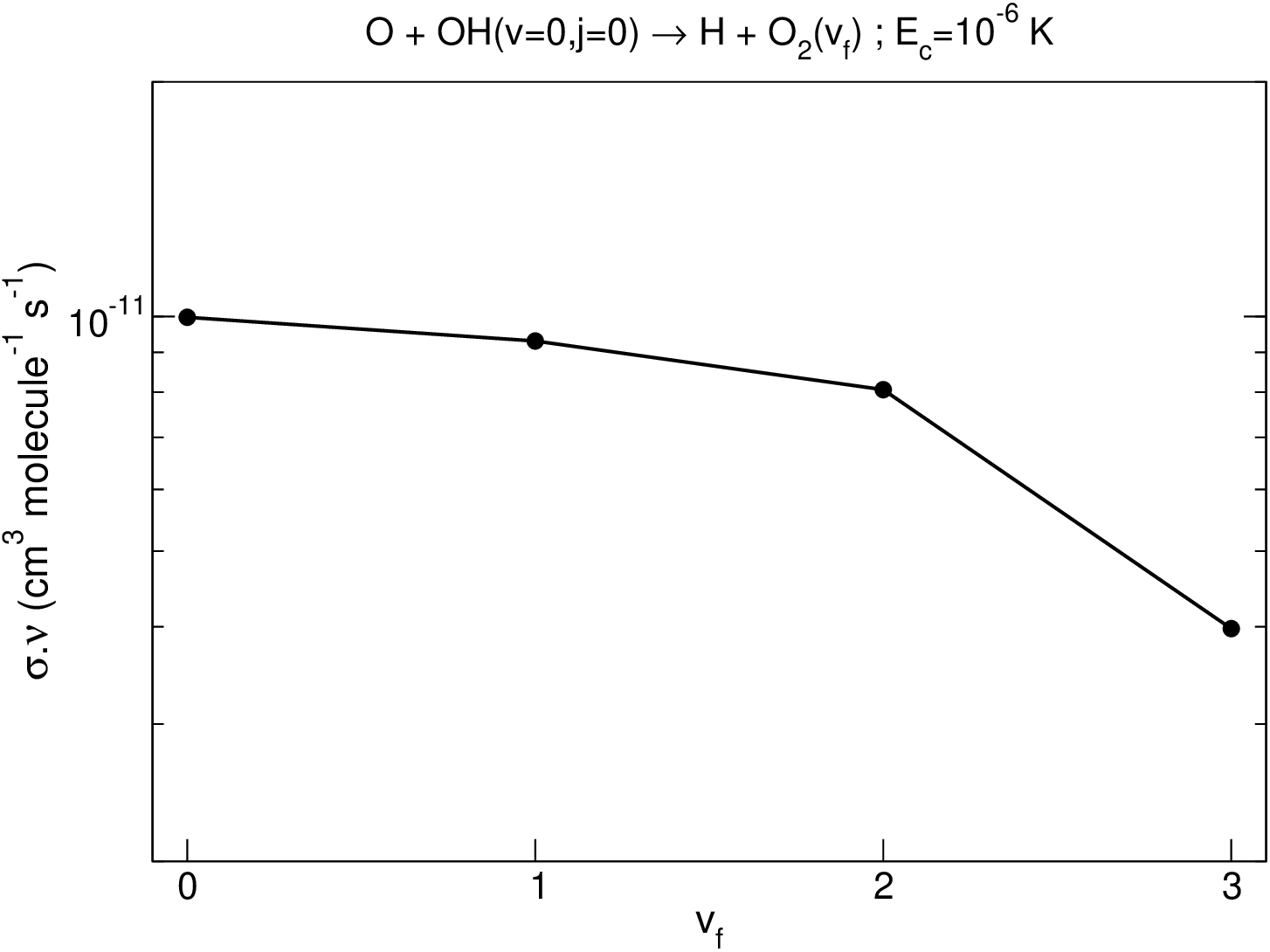}
\caption{
Vibrational populations of the product O$_2$ molecule in
 O + OH$(v=0,j=0)$ $\to$ H + O$_2(v_f)$ reaction evaluated using
the DIMKP PES at a collision energy of 10$^{-6}$~K and $J=0$.
\label{DISTVIB-OHvj00-FIG}
}
\end{center}
\end{figure}
\begin{figure} [h]
\begin{center}
\includegraphics*[width=8cm,keepaspectratio=true,angle=0]{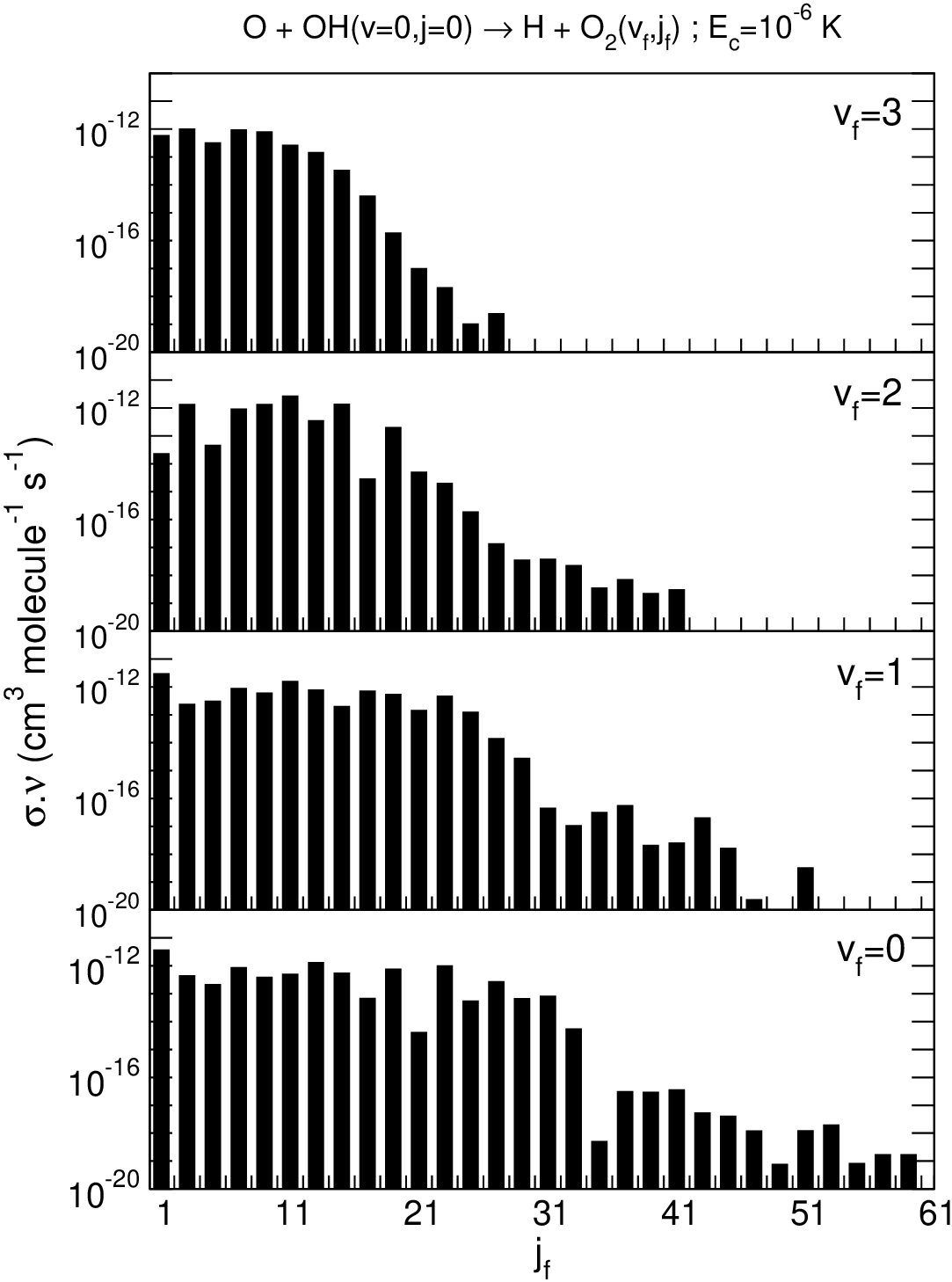}
\caption{
Rotational populations of the product O$_2$ molecule in
O + OH$(v=0,j=0)$ $\to$ H + O$_2(v_f=0,1,2,3,j_f)$ reaction
evaluated using the DIMKP PES at a collision energy of 10$^{-6}$~K and $J=0$.
\label{DISTROT-OHvj00-FIG}
}
\end{center}
\end{figure}

The $J=0$ elastic and reactive cross sections
are plotted as a function of the collision energy in Fig.~\ref{XS-DIMKP-FIG}
for the DIMKP PES.
The corresponding non-thermal rate coefficients
are presented in Fig.~\ref{RATE-DIMKP-FIG}.
In both figures, the Bethe--Wigner laws
are satisfied at ultralow collision energies.
The elastic cross section and the reactive rate coefficient
converge to  values 
of $4.6 \times 10^{-14}$~cm$^2$~molecule$^{-1}$ 
and $3.1 \times 10^{-11}$~cm$^3$~molecule$^{-1}$~s$^{-1}$
respectively
in the limit of zero energy.
The elastic rate coefficient and the reactive cross section behave respectively
as the square root
and the inverse of the square root
of the collision energy.
The Bethe--Wigner regime is reached
at $E_c \approx 10^{-4}$~K, where the reactive rate coefficient is of about one order of magnitude higher
than that of the elastic counterpart.
At $E_c=10^{-6}$~K
the reactive rate coefficient is two orders of magnitude larger
than the  elastic one.
This is reminiscent of other exothermic barrierless systems
such as alkali-metal atom - diatom 
collisions~\cite{Soldan02,Quemener04,Quemener05,Cvitas05a,Cvitas05b,Hutson06,Hutson07a,Quemener07,Cvitas07}.
At energies between $E_c = 0.01$~K and 30~K, elastic and reactive rate coefficients
are of comparable magnitude. 
For $E_c > 30$~K, the elastic scattering is more efficient
than reactive scattering.
The inelastic rotational excitation to the $j=1$ level opens up
at $E_c = 52$~K
and its cross section is
comparable to the reactive contribution.

The $J=0$ contribution to the 
elastic, inelastic and reactive rate coefficients evaluated using the XXZLG PES are 
presented in Fig.~\ref{RATE-XXZLG-FIG}
as functions of the collision energy. 
For $E_c > 1$~K, 
all three rate coefficients
are of comparable magnitudes
with those obtained from the DIMKP PES.
For $E_c < 1$~K,
the reactive rate coefficient is about two orders of magnitude
smaller than that obtained using the DIMKP PES.
The differences can be traced to the incomplete description of the long-range interaction
potential in
the XXZLG PES. 
It does not properly  include the long-range  potential in the O + OH channel
as it was not designed for
quantum dynamics calculations
at ultralow energies. Thus, the comparison of the results obtained using the
two surfaces highlights the crucial role of the long-range interaction potential in
chemical reaction dynamics at low temperatures.
The importance of the long-range intermolecular forces in the O--OH system
in determining its rate coefficient 
has been pointed out by Clary and Werner~\cite{Clary84b}.
In the following, we restrict our calculations to the DIMKP PES.

The vibrational populations 
of molecular oxygen formed in O + OH$(v=0,j=0)$  collisions,
evaluated using the DIMKP PES at an energy
of 10$^{-6}$~K and $J=0$ 
are presented in Fig.~\ref{DISTVIB-OHvj00-FIG}.
Oxygen molecules are predominantly formed
in vibrational levels $v_f=[0-2]$ with a slight preference
for the ground vibrational state $v_f=0$.
Thus, a significant fraction of the O$_2$  molecules formed is
in excited vibrational states.
The rotational level distributions
of molecular oxygen formed in O + OH$(v=0,j=0)$  collisions for $J=0$
are presented in Fig.~\ref{DISTROT-OHvj00-FIG}
for each final vibrational level $v_f$ populated by the reaction at 
an energy of 10$^{-6}$~K.
The results show that low rotational
levels are generally preferred for each of the final vibrational levels. Since the
incident channel includes only the $s$-wave, the final rotational distribution
is largely determined by the anisotropy of the interaction potential.

\begin{figure} [h]
\begin{center}
\includegraphics*[width=8cm,keepaspectratio=true,angle=0]{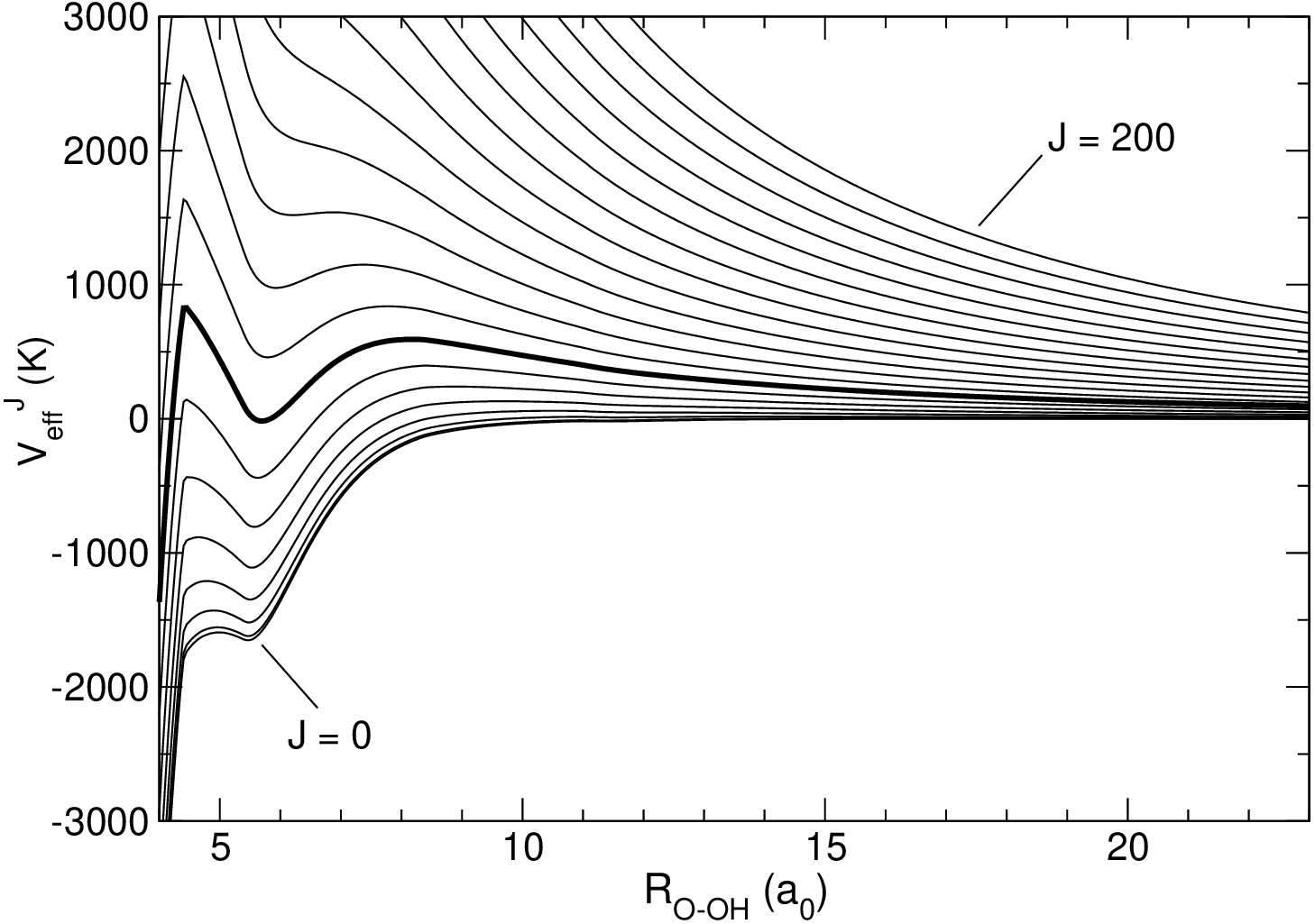}
\caption{
Effective potentials $V^J_{\text{eff}}$ as  functions of $R_{\text{O--OH}}$
for $J=[0-200;10]$ for the DIMKP PES.
The bold line corresponds to $J=70$.
\label{POTEFF-DIMKP-FIG}
}
\end{center}
\end{figure}

Explicit calculations of rate coefficients would require reaction 
probabilities for all contributing values of the total angular momentum quantum number ($J$).
This is a computationally demanding problem for the O + OH reaction if 
full quantum dynamics calculations are employed,
especially when many $J$ values contribute to the reaction probability. The
$J$-shifting approximation~\cite{Bowman91} is widely used to compute rate coefficients when
full quantum calculations are not practical. This is a good approximation
for barrier reactions which involve a transition state but not for 
complex forming reactions. Nevertheless, the $J$-shifting approximation
has been applied to the O + OH reaction in a number of previous 
studies~\cite{Quemener08d,Germann97,Skinner98,Viel98,Xu07}
and it
has been demonstrated that it can predict rate coefficients within about 40\% of
numerically exact calculations~\cite{Lin08}. 
Here we  use the $J$-shifting approximation~\cite{Bowman91}
to compute the rate coefficients for the O + OH reaction.
The  rate coefficient is given by the expression:
\begin{multline}
k_{v,j}(T)=\frac{1}{2 \pi \hbar Q_{\text{R}}}
\times \left( \sum_{J}^{} (2J+1) e^{-E^J_{\text{shift}} / (k_B T)}  \right) \\
\times \int_0^{\infty} P^{\text{re},J=0}_{v,j}(E_c) \ e^{-E_c / (k_B T)} \ dE_c
\label{ratespecified}
\end{multline}
where $k_B$ is the Boltzmann constant, $P^{\text{re},J=0}_{v,j}$ is
the $J=0$ reaction probability and  $E^J_{\text{shift}}$ 
is the height of the effective barrier for a given partial wave $J$
in the entrance channel. To determine the  barrier height
for a given partial wave, we first 
evaluate the effective
potential, $V_{\text{eff}}^J$:
\begin{eqnarray}
V_{\text{eff}}^J=\frac{\hbar^2 J (J+1)}{2 \mu (R_{\text{O--OH}})^2}
+ V_{\text{min}}(R_{\text{O--OH}})
\label{ejshift}
\end{eqnarray}
where $V_{\text{min}}(R_{\text{O--OH}})$
is the minimum energy path of the reaction
as a function of the atom - molecule center-of-mass separation, $R_{\text{O--OH}}$, and
$\mu$ is the reduced mass of the O$-$OH system. 
In Eq.~\eqref{ratespecified},
$Q_{\text{R}} = Q_{\text{trans}} \times  Q_{\text{el}}$ is the reactant partition function. 
For the translational partition function we used the standard formula,
$Q_{\text{trans}}=\left( \frac{\mu k_B T}{2 \pi \hbar^2} \right)^{3/2}$.
For the electronic partition function we used the expression given by Graff and Wagner~\cite{Graff90}:
\begin{eqnarray*}
Q_{\text{el}} =  \frac{(5 + 3 e^{-228/T} + e^{-326/T}) ( 2 + 2 e^{-205/T} )}{2}.
\end{eqnarray*}
The effective barriers, $V^J_{\text{eff}}$, for the DIMKP PES
are shown in Fig.~\ref{POTEFF-DIMKP-FIG}
for  $J = [0 - 200 ; 10]$.
As Fig.~\ref{POTEFF-DIMKP-FIG} illustrates,
the ``reef" visible in the effective potential
at $R_{\text{O--OH}}= 5 \ a_0$
for low values of $J$
becomes an effective barrier
for $J \ge 70$ as indicated by the bold line.
The barrier heights $E^J_{\text{shift}}$
and their locations $R^J_{\text{shift}}$
are reported in table~\ref{TAB3}
for $J = [1 - 10]$.

\begin{figure} [h]
\begin{center}
\includegraphics*[width=8cm,keepaspectratio=true,angle=0]{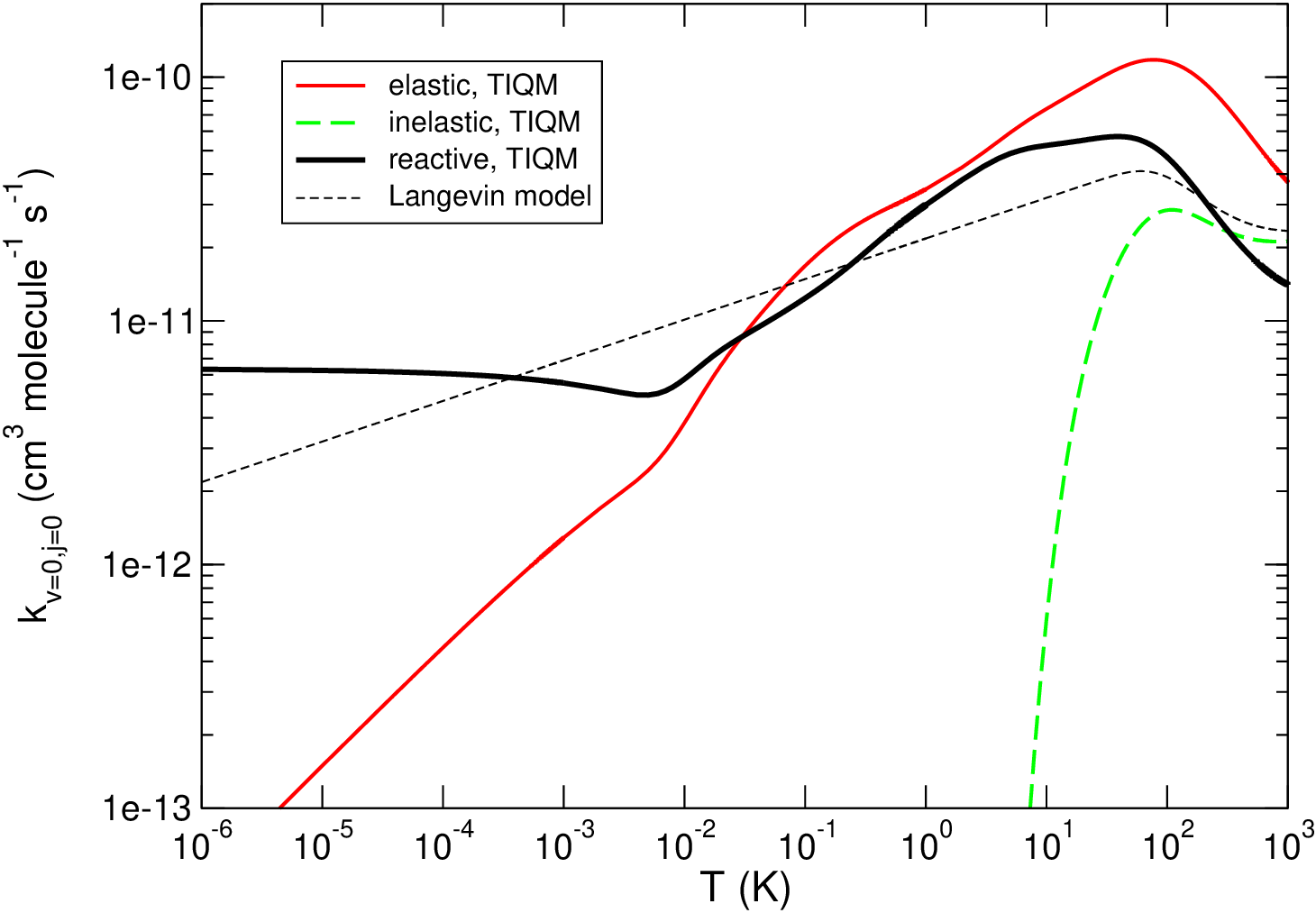}
\caption{
(Color online) 
Rate coefficients of O + OH$(v=0,j=0)$ 
for the elastic (red curve), inelastic (green dashed curve) and reactive (black curve) collisions
for the DIMKP PES. 
The rate coefficient 
from  a classical Langevin model (black dashed curve) is also  shown (see text for detail).
\label{RATE-FIG}
}
\end{center}
\end{figure}

The elastic, inelastic, and reactive rate coefficients of the
O + OH$(v=0,j=0)$ reaction evaluated using the $J$-shifting method
for the DIMKP PES are shown in Fig.~\ref{RATE-FIG}
for $T = 10^{-6} - 10^3$~K.
As indicated earlier, the Bethe--Wigner regime is reached for temperatures
below $T \approx E_c \approx 10^{-4}$~K.
The  value of the rate coefficient in the zero-energy limit 
is $6.2 \times 10^{-12}$~cm$^3$~molecule$^{-1}$~s$^{-1}$.
This is a factor of 5  smaller than the 
rate coefficient reported as $\sigma^{J=0}_{\text{re}} \times \nu$
in Fig.~\ref{RATE-DIMKP-FIG}.
The difference comes from the electronic partition function 
in the denominator in Eq.~\ref{ratespecified},
which is equal to 5 as $T \to 0$.
In  experiments using the stark decelerator methods, OH molecules
were cooled  to $T = 10 - 100$~mK.
In this temperature range, our computed values of 
the elastic rate coefficients are comparable
to the reactive ones 
and will not favor sympathetic cooling of OH by collisions with O atoms. 
Similar conclusions have been found by Lara et al.~\cite{Lara06,Lara07}
for collisions between OH molecules with Rb atoms.
The relatively large rate coefficient for the reaction in the
zero-temperature limit 
indicates that barrierless exothermic reactions occur at significant rates at ultracold temperatures,
in agreement with similar results  for alkali-metal atom - diatom 
reactions~\cite{Soldan02,Quemener04,Quemener05,Cvitas05a,Cvitas05b,Hutson06,Hutson07a,Quemener07,Cvitas07}.
Fig.~\ref{RATE-FIG} shows that the minimum value of the rate  coefficient
for the O + OH reaction is about $4.9 \times 10^{-12}$~cm$^3$~molecule$^{-1}$~s$^{-1}$
at $T=5 \times 10^{-3}$~K.
This provides a lower 
limit for O$_2$ formation
by reaction~\eqref{OpOH}.
We note that the inelastic process becomes more probable than the reactive process
for $T > 330$~K.

\subsection{Classical capture model}

We shall now compare our quantum dynamics 
results with a classical capture model
also known as the Langevin model~\cite{Langevin05}.
This model
has been shown
to work quite well for a certain range of collision energy,
for atom-exchange reactions 
such as K + K$_2$~\cite{Quemener05},
Li + Li$_2$~\cite{Cvitas05a},
and also for non-reactive Rb + OH collisions~\cite{Lara06,Lara07}.
We have shown~\cite{Quemener-CHAPTER} that the Langevin model predicts rate coefficients in
close agreement with experimental values 
for Rb + Rb$_2$~\cite{Wynar00}, Cs + Cs$_2$~\cite{Staanum06,Zahzam06}, 
and Rb/Cs + RbCs~\cite{Hudson08} collisions
reported recently.
For the  O + OH reaction, Clary and Werner~\cite{Clary84b}
used an adiabatic capture theory~\cite{Clary84a},
while
Davidsson and Nyman~\cite{Davidsson88} used a generalized Langevin model.
Both studies showed good agreement between theoretical 
and experimental rate coefficients.
Here we test the validity of the Langevin model 
against a TIQM and $J$-shifting method for the O + OH reaction.
For a non-rotating diatomic molecule ($j=0$)
at large atom - molecule separations $R_{\text{O--OH}}$, for a given value of $J$,
the interaction potential
can be approximated  by
an effective potential, $V_{\text{Lang}}^J$, composed of a
repulsive centrifugal term and an attractive van der Waals term:
\begin{eqnarray*}
V_{\text{Lang}}^J = - \frac{\hbar^2 J (J+1)}{2 \mu (R_{\text{O--OH}})^2}
- \frac{C_6^{\text{O--OH}}}{(R_{\text{O--OH}})^6}.
\end{eqnarray*}
At an atom - diatom distance of
\begin{eqnarray*}
R_{\text{Lang}}^J = \left( \frac{6 \ \mu \ C_6^{\text{O--OH}}}{\hbar^2 \ J (J+1)} \right)^{1/4},
\end{eqnarray*}
the height of the effective potential is given by:
\begin{eqnarray*}
E_{\text{Lang}}^J = \frac{2}{3 \sqrt{3}} \
\ (C_6^{\text{O--OH}})^{-1/2} \ \left(  \frac{\hbar^2 \ J (J+1)}{2 \mu} \right)^{3/2}.
\end{eqnarray*}
The quantities $E_{\text{Lang}}^J$ and $R_{\text{Lang}}^J$ are reported in Tab.~\ref{TAB3}
for $J=1-10$.
The $C_6^{\text{O--OH}}$ coefficient 
is evaluated by fitting the long-range part 
of the minimum energy path $V_{\text{min}}(R_{\text{O--OH}})$.
This yielded a value of $C_6^{\text{O--OH}} \approx 9.2 \ E_h a_0^6$
which is much smaller than the sum of the atom - atom coefficients
$C_6^{\text{O--H}} + C_6^{\text{O--O}} = 24.2 \ E_h a_0^6$. In the DIM model, 
the smaller $C_6^{\text{O--OH}}$ coefficient
is due to the presence of significant diatomic mixing in the OH diatomic
states for $r_{\text{OH}}=1.83 \ a_0$ (see for example Eqs.~$34-37$ 
in Ref.~\cite{Kendrick95}).
This mixing gives rise to an effective $C_6$ coefficient of the
O--HO approach of $\alpha_1 \, C_6^{\text{O--H}} + \alpha_2 \, C_6^{\text{O--O}} = 9.2 \ E_h a_0^6$
where the multiplicative constants $\alpha_1$ and $\alpha_2$
depend on the level of diatomic mixing. As $r_{\text{OH}}$ increases, the diatomic
mixing decreases and $\alpha_1 \to 1$ and $\alpha_2 \to 1$.
The dependence of the long-range coefficients for atom - diatom interactions 
on the diatomic separation has also been noted and investigated in 
recent work on the Li + Li$_2$ and Na + Na$_2$ systems~\cite{jeremy2005,honvault2003}.
For the atom - molecule reduced mass we used  $\mu = 15023.74 \ au$.
In the Langevin model, the rate coefficient as a function of the temperature is given by the expression:
\begin{eqnarray*}
k_{\text{Lang}}(T)= \frac{\pi}{Q_{\text{el}}} \left( \frac{8 k_B T}{\pi \mu} \right)^{1/2}
\left( \frac{2 \, C_6^{\text{O--OH}}}{k_B T} \right)^{1/3} \Gamma(2/3).
\end{eqnarray*}

\begin{table}[h]
\begin{center}
\begin{tabular}{|c | c  c | c c|}
\hline
$J$ & $E^J_{\text{shift}}$ (K)  & $R^J_{\text{shift}}$ (a$_0$) & $E_{\text{Lang}}^J$ (K) & $R_{\text{Lang}}^J$ (a$_0$) \\ [0.5ex]
\hline
1 & 0.022 & 25.4 & 0.022 & 25.4 \\
2 & 0.11 & 19.3 & 0.11 & 19.3 \\
3 & 0.30 & 18.6 & 0.32 & 16.3 \\
4 & 0.54 & 18.4 & 0.68 & 14.3 \\
5 & 0.86 & 18.0 & 1.26 & 12.9 \\
6 & 1.26 & 17.5 & 2.08 & 11.9 \\
7 & 1.75 & 17.1 & 3.20 & 11.1 \\
8 & 2.34 & 16.7 & 4.68 & 10.4 \\
9 & 3.04 & 16.4 & 6.53 & 9.8 \\
10 & 3.84 & 16.1 & 8.83 & 9.3 \\ [0.5ex] 
\hline
\end{tabular}
\end{center} 
\caption{Comparison of
the heights and positions
of the effective barriers
between the $J$-shifting approximation
and the Langevin model.
\label{TAB3}
}
\end{table}

The rate coefficient predicted by the Langevin model is plotted
in  Fig.~\ref{RATE-FIG} (black dashed curve).
Except in the Bethe--Wigner regime for $T < 10^{-4}$~K, 
where the classical model is not valid, 
the Langevin model predicts rate coefficients in semi-quantitative agreement
with those obtained from the quantum calculations.
Overall, the reactive rate coefficients oscillate slightly around the Langevin line,
as in other barrierless systems
such as K + K$_2$~\cite{Quemener05} and Li + Li$_2$~\cite{Cvitas05a}.
However, the Langevin model yields good agreement
only for a restricted range of temperatures.
The lower limit of the model prediction is restricted by the number
of partial waves included in the calculations.
In previous studies~\cite{Quemener05,Cvitas05a}, it has been found that
when
three or more partial waves are included,
the quantum and classical results are in good agreement.
If less than three partial waves are involved,
the quantum character becomes dominant and the results cannot be 
compared
with the classical model.
These previous studies have also shown 
that the maximum of
the quenching rate coefficient for a given partial wave $J$
occurs at about a collision energy
comparable to 
the height of the barrier, $E^J_{\text{Lang}}$.
Thus, the lower limit
corresponds to a collision energy 
of $E^{J=3}_{\text{Lang}} = 0.32$~K (see Tab.\ref{TAB3})
for the present system.
The upper limit
is bounded by the long-range part of the potential.
When the Langevin radius $R^J_{\text{Lang}}$
is located at a distance where the long-range part of the potential
is not described 
by the van der Waals interaction,
$E^J_{\text{Lang}}$
will differ from
$E^J_{\text{shift}}$.
In this case, the classical results
will differ from the quantum calculations.
Tab.~\ref{TAB3} shows that
this is the case
for $J=3$ and the upper limit of the model
also corresponds to a collision energy of $E^{J=3}_{\text{Lang}} = 0.32$~K.
Thus, the 
Langevin model gives quantitative agreement
for temperatures around $T \approx 0.3$~K,
as seen in Fig.~\ref{RATE-FIG}.
For $T > 0.3$~K,
the rate coefficient depends on the exact details of the effective potential
in the entrance channel.
However, the classical result
is in overall
agreement
with the quantum result, though the classical model
predicts a smaller value for the rate coefficient.
This is because  $E^J_{\text{shift}}$
$<$ $E^J_{\text{Lang}}$ for $J > 3$
(see Tab.~\ref{TAB3})
and reactivity is less probable to occur with 
the Langevin model.
Therefore, the Langevin model 
provides a lower limit of the rate coefficients
for $T = 1 - 100$~K which affirms
the theoretical conclusions of Ref.~\cite{Quemener08d}
and experimental conclusions of Ref.~\cite{Carty06},
that the rate coefficient of reaction~\eqref{OpOH}
is unlikely to vanish for $T<10$~K.
We also note that the Langevin rate coefficient is in semi-quantitative 
agreement with the inelastic rate coefficient
in the range $T=100-1000$~K. The difference is less than 30~\%.

\section{Conclusion}

This paper 
presents the first quantum mechanical investigation
of an ultracold barrierless reaction
with chemically distinct reactants and products.
We investigated dynamics of
molecular oxygen formation in ultracold collisions of the hydroxyl radical
and atomic oxygen using a time-independent quantum mechanical method
based on hyperspherical coordinates.
It has been found that formation of molecular oxygen
occurs with a relatively large  rate coefficient
of $6.2 \times 10^{-12}$~cm$^3$~molecule$^{-1}$~s$^{-1}$
at ultracold temperatures.
The oxygen molecules are mainly formed
in the $v=0-2$ vibrational levels
with a slight preference
for the $v=0$ level.
Calculations show that at temperatures of $T = 10 - 100$~mK,
the elastic cross sections
are not large enough
to achieve efficient evaporative cooling in collisions
between OH molecules and O atoms.
We predict a lower limit 
of $4.9 \times 10^{-12}$~cm$^3$~molecule$^{-1}$~s$^{-1}$
for the
rate coefficient for the O + OH($v=0,j=0$) reaction at $T=5 \times 10^{-3}$~K.
This shows that formation of O$_2$ molecules is significant even
at ultracold temperatures.
It has been found that
a classical capture model
is valid
for temperatures around $T \approx 0.3$~K
where quantum and classical calculations
yield comparable results.
Based on our analysis of the long-range interaction potential
we find that for $T = 1 - 10$~K
the Langevin model can provide
a lower limit
to the quantum reactive rate coefficients
calculated 
within the $J$-shifting
approximation.

Future work will consider the effects of the geometric phase and non-adiabatic
couplings between different PESs of HO$_2$ on the reaction dynamics. The
geometric phase due to the conical intersection of two PESs may have an
important effect at low and ultralow collision energies where only a few
partial waves contribute to the reaction probabilities.

\section{Acknowledgments}

This work was supported  by NSF grants \#PHY-0555565 (N.B.) and \#ATM-0635715 (N.B.).
B.K.K. acknowledges that part of this work was done under the auspices of the US
Department of Energy at Los Alamos National Laboratory. Los Alamos National
Laboratory is operated by Los Alamos National Security,
LLC, for the National Nuclear Security Administration of the
US Department of Energy under contract DE-AC52-06NA25396.


\begin{thebibliography}{46}
\bibitem{NOTE}
In our previous study~\cite{Quemener08d}
we also used $\rho_{\text m}=44.0$~a$_0$
for collision energies $E_c < 10^{-4}$~eV.
\end{thebibliography}
\end{document}